\documentclass{article}
\usepackage{spconf,amsmath,graphicx}
\usepackage{multirow}
\usepackage{float}
\usepackage{enumitem}
\usepackage[table,xcdraw]{xcolor}

\usepackage[normalem]{ulem}
\usepackage{hyperref}
\usepackage{makecell}
\usepackage{bibspacing}
\usepackage{enumitem}
\useunder{\uline}{\ul}{}

\makeatletter
\renewcommand\section{\@startsection {section}{1}{\z@}%
                                    {-2.5ex \@plus -1.5ex \@minus -.05ex}%
                                    {1.5ex \@plus0.5ex}%
                                    {\normalfont\Large\bfseries}}
\renewcommand\subsection{\@startsection{subsection}{2}{\z@}%
                                      {-2.8ex\@plus -1.5ex \@minus -.05ex}%
                                      {1.0ex \@plus 0.12ex}%
                                      {\normalfont\large\bfseries}}
                
\renewcommand\subsubsection{\@startsection{subsubsection}{3}{\z@}%
                                          {-1.2ex\@plus -0.5ex \@minus -.2ex}%
                                          {1.0ex \@plus .12ex}%
                                          {\normalfont\normalsize\bfseries}}

\newcommand{\taejin}[1]{\textcolor{black}{#1}}
\usepackage{fancyhdr}
\fancypagestyle{pageStyleOne}{%
    
    \fancyhf{}
    \fancyfoot[L]{\scriptsize © 20XX IEEE. Personal use of this material is permitted. Permission from IEEE must be obtained for all other uses, in any current or future media, including reprinting/republishing this material for advertising or promotional purposes, creating new collective works, for resale or redistribution to servers or lists, or reuse of any copyrighted component of this work in other works.}
}

\title{TitaNet: Neural Model for speaker representation with 1D Depth-wise separable convolutions and global context }
\name{Nithin Rao Koluguri, Taejin Park, Boris Ginsburg}
\address{NVIDIA, USA}
\begin{document}
%
\maketitle
\thispagestyle{pageStyleOne}

\begin{abstract}

In this paper, we propose TitaNet, a novel neural network architecture for extracting speaker representations. 
We employ 1D depth-wise separable convolutions with Squeeze-and-Excitation (SE) layers with global context followed by channel attention based statistics pooling layer to map variable-length utterances to a fixed-length embedding (t-vector). TitaNet is a scalable architecture and achieves state-of-the-art performance on speaker verification task with an \taejin{equal error rate (EER) of} 0.68\% on the VoxCeleb1 trial file and also on speaker diarization tasks with diarization error rate (DER) of 1.73\% on AMI-MixHeadset, 1.99\% on AMI-Lapel and 1.11\% on CH109. Furthermore, we investigate various sizes of TitaNet and present a light TitaNet-S model with only 6M parameters that achieve near state-of-the-art results in diarization tasks.

\end{abstract}
\section{Introduction}
\label{sec:intro}

Speaker recognition is a broad research area that solves two major tasks based on characteristics of voices: speaker identification and speaker verification.
Speaker identification is about identifying a person and speaker verification is about verifying whether the speaker is who they claim to be. 
Speaker diarization is a task of partitioning audio recordings into speaker-homogeneous segments belonging to each individual speaker. 
Typically, speaker recognition and  diarization systems operate on unconstrained speech utterances, which are converted to a vector of fixed length, called speaker embeddings. These speaker embeddings represent the identity of each speaker and are used for speaker recognition and speaker diarization tasks.

In recent years, deep neural networks (DNNs) have been actively employed for speaker embedding extractors since d-vector \cite{Variani2014} was proposed. Subsequently, x-vector~\cite{snyder2018} was widely used because of the superior performance achieved by employing statistical pooling and time delay neural network (TDNN). Other architectures such as ResNet-based convolutional neural networks (CNNs) \cite{yu2019ensemble} and  CNNs with cross convolutional layers \cite{gao2018improved} were employed for capturing the traits of speech. In addition, to cope with the variable-length inputs, Transformer \cite{safari2020self}, CNN-LSTM \cite{jung2018complete} and a slew of variants of TDNN \cite{villalba2019state, zhu2020orthogonal, dawalatabad2021ecapa} were applied for DNN-based speaker embedding extractors.   

In this work, we develop a speaker embedding extractor model that shows superior performance on both speaker verification and diarization tasks. We adopt the architecture from  ContextNet~\cite{han2020contextnet},  a state-of-the-art  automatic speech recognition (ASR) model, which combines local features from 1D depth-wise separable convolutions and global context from Squeeze and Excitation layers. 

We train text-independent speaker recognition models where the identity of the speaker is based on how speech is spoken, not necessarily on what is being said.
The following are the main contributions of this paper:
\begin{itemize}[leftmargin=2.0ex,itemsep=0.2pt,topsep=0pt]
    \item We propose use of 1D separable depth-wise convolutions compared to full 1D convolutions. 
    \item We bring in global context to speaker embedding models by introducing global average pooling after the squeeze and excitation module. This contrasts with the embedding extractors based on TDNN, such as x-vector~\cite{snyder2018} or most recently ECAPA-TDNN~\cite{desplanques2020ecapa}.
    
    \item TitaNet-M is half the size of comparable speaker embedding extractors like ECAPA-TDNN or Conformer-based baselines and achieves superior performance in speaker diarization.
    
    \item We train our networks end-to-end using angular softmax margin loss and use cosine similarity as a backend for speaker representations. Such approach leads us to avoid the burden of training external models like probabilistic linear  discriminant analysis  (PLDA)~\cite{kenny2010bayesian} and agglomerative hierarchical clustering (AHC)~\cite{sell2018diarization} as in well-known speaker verification systems \cite{snyder2018, villalba2019state}, or speaker diarization systems~\cite{sell2018diarization, landini2022bayesian}.
 
\end{itemize}

We introduce three models with different sizes. The architecture is easily scalable in both depth and width by design, and we show that the scaling in width is very effective in reducing the model size with a small change in performance. In the experimental section, we demonstrate the performance of the model on a VoxCeleb1 cleaned test speaker verification trial file. In addition, we evaluate diarization performance on popular evaluation datasets like AMI (Lapel), AMI (MixHeadset) ~\cite{carletta2005ami}, NIST-SRE-2000~\cite{martin2001nist} and CH109~\cite{canavan1997callhome}.

\section{Model Architecture}
\begin{figure}[t]
    \centering
    \includegraphics[width=8.5cm]{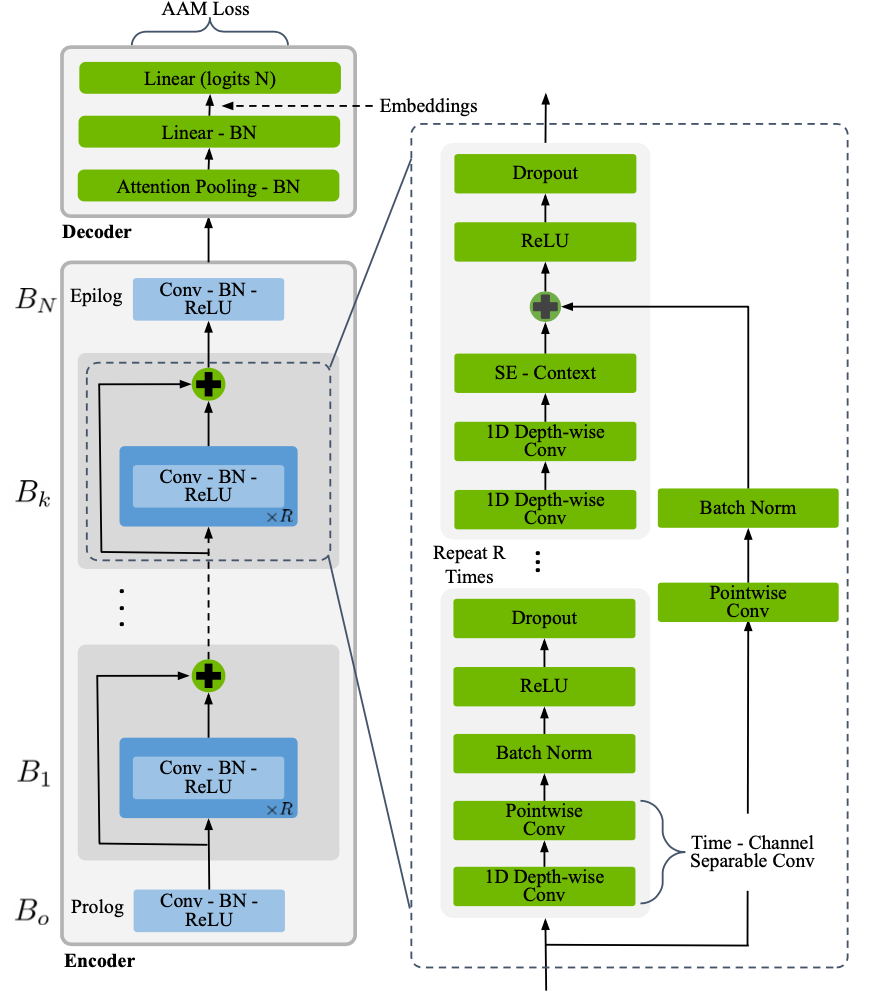}
    \caption{TitaNet Encoder and Decoder Architecture}
    \label{fig:model_figure}
\end{figure}

\subsection{Encoder}
\label{subsec:encoder}
The model is based on the ContextNet ASR architecture \cite{han2020contextnet} comprising of an encoder and decoder structure. We use the encoder of the ContextNet model as a top-level feature extractor, and feed the output to the attentive pooling layer. This layer computes attention features across channel dimensions to capture the time-independent utterance-level speaker representations.

TitaNet is a 1D time depth-wise channel separable convolutional model with ContextNet-like architecture combined with channel attention pooling. Fig.~\ref{fig:model_figure} describes the ContextNet-$B$$\times$$R$$\times$$C$ model encoder and attention pooling decoder, where $B$ is the number of blocks, $R$ is the number of repeated sub-blocks per block, and $C$ is the number of filters in the convolution layers of each block. The encoder starts with a prologue block $B_{0}$, followed by mega blocks $B_{1}$ $\dots$ $B_{N-1}$ and ends with epilogue block $B_{N}$. Prologue and epilogue blocks differ from mega blocks, they both have the same convolution module (Conv), batchnorm and relu layers and have fixed kernel sizes of 3 in prologue and 1 in epilogue for all the network architectures we propose. They do not contain residual connections and dropout layers. Each mega block begins with time-channel separable convolutional\cite{kriman2020quartznet} layer with stride 1  and dilation 1, followed by batchnorm, relu and dropout.  

Each time-channel separable convolution module is made up of two parts: a depth-wise convolutional layer and a pointwise convolutional layer. Depth-wise convolutions apply a single filter per input channel (input depth). Pointwise convolutions are $1$$\times$$1$ convolutions, used to create a linear combination of the outputs of the depth-wise layer. These layers are repeated $R$ times, which can be modified to vary the depth of the network. These repeated layers are residually connected with Squeeze and Excitation layers with global average pooling for context inclusion. By using global context, the SE layer squeezes a sequence of local feature vectors into a single global context vector, broadcasts this context back to each local feature vector, and merges the two via multiplications. The width of the network can be increased or decreased by varying output channel filter sizes of each mega block. For TitaNet models, width and depth are changed by varying these filter sizes, $C$ and the number of repeated layers, $R$ respectively.

\subsection{Decoder and Embeddings}
The top level acoustic features obtained from the output of encoder are used to compute intermediate features that are passed to the decoder for getting utterance level speaker embeddings. The intermediate time-independent features are computed using an attentive statistics pooling layer \cite{dawalatabad2021ecapa}, where the channel attention features $E$ are computed across time-channels to get a time-independent feature representation $S$ of size $B$$\times$$3072$. 

The intermediate features $S$ are passed through the Decoder consisting of two linear layers, one of output size 192 and another for a linear transformation from 192 to the final number of classes $N$, to compute the probability that the current segment belongs to a speaker from the training set. In this fashion, the network extracts fixed-length representation from variable length speech segments. We extract t-vectors before the final logits linear layer of fixed size $192$. 
\subsection{Loss function}
The TitaNet model was trained end-to-end with additive angular margin (AAM) loss~\cite{deng2019arcface}. The AAM  helps to optimize the cosine distance between speaker embeddings. For all the verification and diarization experiments presented in this paper, we use cosine similarity as the back-end:
\begin{equation}
\mathcal{L} = -\frac{1}{N} \sum_{i=1}^{N} \log \frac{e^{s\left(\cos \left(\theta_{y_{i}}+m\right)\right)}}{e^{s\left(\cos \left(\theta_{y_{i}}+m\right)\right)}+\sum_{j=1, j \neq y_{i}}^{n} e^{s \cos \theta_{j}}}
\end{equation}
where $m$ is margin, $s$ is scale and $\theta_{j}$ is the angle between the final linear layer weight $W_{j}$ and incoming feature $x_{i}$. Here $m$ and $s$ are predefined hyper parameters. 

\section{Experiments}
\label{sec:experiments}

We designed three TitaNet models: TitaNet-S with 256 channels, TitaNet-M with 512 channels, and  
TitaNet-L with 1024 channels. All models have the same number of repeating layers $R$ as 3, and the same kernel filter sizes as 3,7,11 and 15.  TitaNet-S has 6.4M , TitaNet-M -- 13.4M, and TitaNet-L -- 25.3M parameters.

\subsection{Datasets}
\subsubsection{Training Data}
We use following datasets to datasets to train TitaNet: VoxCeleb1 and VoxCeleb2 dev\cite{Chung2018}, NIST SRE portion of datasets from 2004-2008 (LDC2009E100), Switchboard-Cellular1 and Switchboard-Cellular2~\cite{godfrey1993switchboard}, Fisher~\cite{cieri2004fisher},  and Librispeech~\cite{panayotov2015librispeech}  (see Table~\ref{tab:datasets}). 
Combined, these datasets consist of about 4.8M utterances from 16.6K speakers. 
We augment the training data with RIR~\cite{ko2017study} impulse corpora, speed pertubation with 0.95x \& 1.05x speeds and also spec augment~\cite{park2019specaugment}. 

\subsubsection{Evaluation Data}
We use the VoxCeleb1 cleaned test trial file to evaluate EER for speaker verification. We use the following three datasets for the evaluating speaker diarization system:  
\begin{itemize}[leftmargin=2.0ex,noitemsep,topsep=0pt]
\item  NIST-SRE-2000~\cite{martin2001nist}: all sessions from LDC2001S97.
\item AMI Corpus~\cite{carletta2005ami}:  Lapel and MixHeadset audio subsets from partition set \cite{Bredin2020}.
\item CH109~\cite{canavan1997callhome}: we use a subset of CALLHOME American English speech (CHAES), which contains only two speakers. There are 109 sessions in this subset. The remaining 11 sessions in CHAES are used as a dev set for CH109 and NIST-SRE-2000.
\end{itemize}

\begin{table}[t]
\centering
\resizebox{\columnwidth}{!}{%
\begin{tabular}{|c|c|c|c|}
\hline
\textbf{Dataset}     & \textbf{\# of Speakers} & \textbf{\begin{tabular}[c]{@{}c@{}}Duration\\ (in Hrs)\end{tabular}} & 
\textbf{\begin{tabular}[c]{@{}c@{}}\# Utterances\\ (in K)\end{tabular}} \\ \hline \hline
VoxCeleb1   & 1211           & 227                                                     & 332                                                         \\ 
VoxCeleb2   & 5994           & 1895                                                   & 2274                                                        \\ 
SRE         & 3787           & 503                                                    & 944                                                         \\ 
Fisher      & 951            & 162                                                    & 278                                                       \\ 
Switchboard & 2400           & 247                                                     & 425                                                         \\ 
LibriSpeech & 2338           & 336                                                     & 634                                                         \\ 
Total       & 16681          & 3373                                                     & 4890                                                        \\ \hline
\end{tabular}%
}
\caption{Statistics of each dataset used for training TitaNet}
\label{tab:datasets}
\end{table}

\begin{figure}[t]
    \centering
    \includegraphics[width=0.9\columnwidth]{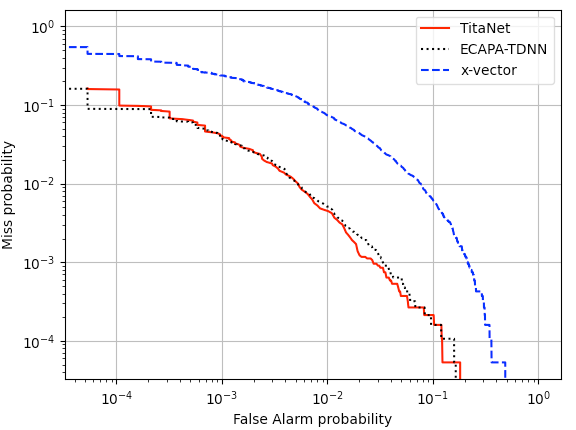}
    \vspace{-3ex}
    \caption{DET curve for VoxCeleb1 cleaned trial comparing with previous studies}
    \label{fig:der_curve}
\end{figure}

\begin{table}[b]
\centering
\resizebox{\columnwidth}{!}{%
\begin{tabular}{|c|c|c|c|}
\hline
\multirow{2}{*}{\textbf{Models (Backend)}} & \multirow{2}{*}{\textbf{\begin{tabular}[c]{@{}c@{}}\# Params\\ (M)\end{tabular}}} & \multicolumn{2}{c|}{\textbf{VoxCeleb1}} \\ 
\cline{3-4} 
  &   & \textbf{EER (\%)}   & \textbf{MinDCF} 
\\ 
\hline \hline
x-vector (PLDA) \cite{snyder2018} & 9 & 2.97  & 0.323 \\ 
ECAPA (CS) \cite{dawalatabad2021ecapa}    & 22.3  & 0.69    & \textbf{0.082}        
\\ Conformer (CS)\cite{gulati2020conformer}             &   26.4                & 2.43                  & 0.264        \\ 
\hline 
TitaNet-S  (CS)    & 6.4    &  1.15 &  0.131 \\ 
TitaNet-M  (CS)    & 13.4   & 0.81  & 0.106  \\ 
TitaNet-L  (CS)  & 25.3    & \textbf{0.68} & 0.087  \\ \hline 
\end{tabular}%
}
\caption{TitaNet comparison with other models for speaker verification task. All models has been evaluated with Cosine Similarity (CS) backend except x-vector which used PLDA.}

\label{tab:TitaNet_models}
\end{table}

\subsection{Experiment Setup}
Every speaker recognition experiment consists of common data pre-processing steps for training, development, and evaluation steps. During the pre-processing, we do not use a speech activity detector (SAD) to avoid dependence on an additional model. Instead, we split speech segments longer than 3 sec into random chunks of 1.5, 2, and 3 sec. We compute acoustic features for every 25 ms frame window shifted over 10 ms. The acoustic features are 80-dimensional mel spectrograms  computed using a 512 FFT and a Hann window. Next, Mel-spectrogram features are  normalized over the frequency axis. Every utterance fed to the encoder has size of $T$$\times$$80$, where $T$ is the number of frames in a given speech utterance file. The accuracy of speaker verification systems is measured using EER and minimum normalized detection cost (MinDCF) with $P_{target}$ = $10^{-2}$ and $C_{FA}$ = $C_{Miss}$ = $1$. Both the EER for verification and the DER - for diarization are done using cosine similarity (CS) back-end. 

In our diarization experiments, the evaluation datasets are divided into two groups: telephonic and non-telephonic speech. Based on experiments with the dev sets, we found window size of 1.5 sec with a shift of 0.75 sec works best for telephonic speech. For non-telephonic speech, the best settings were 3 sec and 1.75 sec for window and shift respectively. In the evaluation datasets used in this paper, AMI Lapel and MixHeadset fall under non-telephonic speech, and the rest of them are in the telephonic speech group. Unlike the previous studies~\cite{dawalatabad2021ecapa, sell2018diarization}, we do not use external dev data to tune the clustering parameters by relying on an auto-tuning approach~\cite{park2019auto}. Similar to the previous systems~\cite{dawalatabad2021ecapa}, we use collar 0.25 sec and ignore overlap speech regions for speaker error rate calculation. All TitaNet models in Table \ref{tab:TitaNet_models} are trained for 250 epochs with SGD optimizer, with initial learning rate (LR) $0.08$ using cosine annealing LR scheduler on 4 nodes with 8 V100 GPUs per node.

\subsection{\taejin{Evaluation Results}}

\begin{table}[t]
\centering
\resizebox{\columnwidth}{!}
{
\begin{tabular}{|c|c|c|c|c|}
\hline
\textbf{Models (Backend)} & \textbf{\begin{tabular}[c]{@{}c@{}}NIST-SRE\\ 2000\end{tabular}} & \textbf{\begin{tabular}[c]{@{}c@{}}AMI\\ Lapel\end{tabular}} & \textbf{\begin{tabular}[c]{@{}c@{}}AMI\\ MixHeadset\end{tabular}} & \textbf{CH109} \\ \hline \hline
 x-vector (PLDA + AHC) \cite{snyder2018}  & 8.39  & - & - & 9.72   \\
 ECAPA (SC) \cite{dawalatabad2021ecapa} &
 -   & 2.36  & 1.78 & -   \\
 x-vector (MCGAN) \cite{pal2021meta}& \textbf{5.73}  & - & - & -   \\
 \hline
 TitaNet-S (NME-SC)   &  6.37 &  2.00  &  2.22   &   \textbf{1.11} \\
 TitaNet-M (NME-SC) & 6.47 & \textbf{1.99} & 1.79  & 1.13  \\
 TitaNet-L (NME-SC) & 6.73 & 2.03 & \textbf{1.73}  & 1.19  \\
\hline
\end{tabular}
}
\label{tab:known_num_of_speakers}
\caption{TitaNet comparison with other models for speaker diarization with oracle SAD known speakers number, DER($\%$).
}
\vspace{-3ex}
\end{table}

\subsubsection{Speaker Verification}
In the speaker verification experiments, we train the model on the datasets shown in the Table~\ref{tab:datasets}. We train these systems initially as a speaker identification model with 10 percent of audio files of each speaker set aside as validation  data from training sets. With this setup, we trained TitaNet models end-to-end using additive margin angular loss.
Table \ref{tab:TitaNet_models} shows the performance of TitaNet models on the VoxCeleb cleaned trial file.
We observed a \taejin{high-degree} of  sensitivity  on  validation  curves  with  slight  variations in the margin ($m$) and scale ($s$) for angular loss. With $m$ as 30 and $s$ as 0.2 TitaNet-L showed state of the art performance with EER of 0.68\% on VoxCeleb1 cleaned test trial file outperforming previously reported results \taejin{in} \cite{dawalatabad2021ecapa,zeinali2019description}.

As it can be noticed from Table \ref{tab:TitaNet_models}, Titanet models are easily scalable to achieve very competitive performances even with relatively few parameters around 6M. These models show a direct relationship on accuracy in contrast to the number of parameters. We show the Detection Error Trade-off (DET) curves to compare TitaNet-L model with other previously stated CNN based models. 

\subsubsection{Speaker Diarization}
We employ our proposed speaker embedding extractor models  for speaker diarization tasks. Cosine similarity is used for measuring the distance between speaker embeddings and Normalized Maximum Eigengap Spectral Clustering (NME-SC) \cite{park2019auto} on the extracted embeddings to obtain the clustering result. We show the performance of each TitaNet model on popular evaluation datasets as shown in Tables 3 and 4.
The diarization experiments are based on oracle SAD to evaluate the SAD-independent performance. In Table 3, we show the results for known number of speakers case and in Table 4, we present the results for unknown number of speakers for which the speaker count is estimated using NME-SC clustering algorithm. 

TitaNet models outperform the previous state-of-the-art models on the AMI-Lapel, AMI-MixHeadset and CH109 evaluation datasets. It is worth noting that the performance of the small and medium TitaNet models show minor differences even if we reduce their model parameters by $2x$ and $4x$ respectively, compared to the largest model. 
In diarization systems we believe there is no major performance improvement using larger TitaNet models when compared to verification. We hypothesize that this is related to the fact that separability in the embedding space does not require a higher level of precision since the clustering process only involves relatively few speakers.
The TitaNet models show very good improvement on all datasets,  except on the NIST-SRE-2000 evaluation set. But note that the clustering approaches in~\cite{landini2022bayesian, snyder2018} involves additional training for Hidden Markov Model or PLDA.

\begin{table}[t]
\centering
\resizebox{\columnwidth}{!}
{
\begin{tabular}{|c|c|c|c|c|}
\hline
\textbf{Models (Backend)} & \textbf{\begin{tabular}[c]{@{}c@{}}NIST-SRE\\ 2000\end{tabular}} & \textbf{\begin{tabular}[c]{@{}c@{}}AMI\\ Lapel\end{tabular}} & \textbf{\begin{tabular}[c]{@{}c@{}}AMI\\ MixHeadset\end{tabular}} & \textbf{CH109} \\ \hline \hline
x-vector (PLDA + AHC) \cite{snyder2018}  & 7.12  & - & - & -   \\
x-vector (VBx) \cite{landini2022bayesian}
& \textbf{4.42} & - & 2.17 & - \\
ECAPA (SC)\cite{dawalatabad2021ecapa} 
& - & 2.13 & 2.17 & -   \\
x-vector (MCGAN) \cite{pal2021meta} 
& 6.76 & - & $4.92^*$ & -   \\
\hline
TitaNet-S (NME-SC) & 5.49 & 2.3 & 1.97 & \textbf{1.42} \\ 
TitaNet-M (NME-SC) & 5.75  & 2.59 & 1.89 & 1.51 \\
TitaNet-L  (NME-SC) & 5.38 & \textbf{2.03} & \textbf{1.89}   & 1.63  \\
\hline
\end{tabular}
}
\smallskip
\parbox[t]{\textwidth}{\footnotesize
  {$*$This number is reported based on Kaldi evaluation partition~\cite{landini2022bayesian}}
}
\label{tab:unknown_num_speakers}
\caption{TitaNet comparison with other models for speaker diarization with oracle SAD estimated number of speakers, DER($\%$).}
\vspace{-3ex}
\end{table}

\section{CONCLUSION}

In this paper, we present TitaNet, a new speaker representation learning model that utilizes the global context of squeeze-and-excitation layers combined with channel attention pooling for extracting fixed length speaker embeddings. The model employs 1D depth-wise separable convolutions for speaker embedding models that showed state-of-the-art performance in ASR tasks. The TitaNet-M model, which is half the size of previous state-of-the-art systems outperforms them in speaker diarization tasks while achieving competitive numbers on verification tasks. The TitaNet-L model significantly outperforms existing models in speaker verification and diarization tasks. 

The models' implementation and pre-trained checkpoints are made available through NVIDIA NeMo toolkit \cite{kuchaiev2019nemo}.\footnote{https://github.com/NVIDIA/NeMo} 

\section{Acknowledgments}
We would like to thank NVIDIA AI Applications team for the help and valuable feedback.


\label{sec:refs}
\bibliographystyle{IEEEbib}
\bibliography{strings,speakerbib}

\end{document}